\documentclass[a4paper,twocolumn,amsmath,amssymb,aps,prl,10pt,superscriptaddress,showpacs,groupedaddress]{revtex4-1}
\usepackage{amsmath}
\usepackage{amssymb}
\usepackage{graphicx}
\usepackage[utf8]{inputenc}
\usepackage[T1]{fontenc}
\usepackage{color}
\newcommand{\av}[1]{\langle\, #1\,\rangle}
\renewcommand{\i}{\mathrm{i}}
\usepackage{braket}
\begin{document}
\title{Scaling up the Anderson transition in random-regular graphs}
\author{M.\ Pino}
\affiliation{Institute of Fundamental Physics IFF-CSIC, Calle Serrano 113b, Madrid 28006, Spain}

\begin{abstract}
We study the Anderson transition in lattices with the connectivity of a random-regular graph. Our results indicate that fractal dimensions are continuous across the transition, but a discontinuity occurs in their derivatives, implying the non-ergodicity of the metal near the Anderson transition. A critical exponent $\nu = 1.00 \pm0.02$ and critical disorder $W= 18.2\pm 0.1$ are found via a scaling approach. Our data support that the predictions of the relevant Gaussian Ensemble are only recovered at zero disorder.
\end{abstract}
\maketitle

\paragraph{Introduction.---}  Anderson localization of a single-particle  is crucial to understand transport in disordered materials\ \cite{An1958}. However, interactions between localized states have an important effect on those materials as, for instance, they affect the phonon-assisted conduction at low temperature\  \cite{efros1975coulomb,kim1993observation,rosenbaum1991crossover,pollak2013electron,pino2012coulomb}. Fleishman and Anderson argued that the Anderson transition, thus localization, survives upon the inclusion of short-range interactions in three dimensions\ \cite{fleishman1980interactions}.  Altshuler and coworkers found later a metal-insulator transition, the many-body localization transition, for disordered and interacting one-dimensional systems where all single-particle states are localized\ \cite{Basko2006}. Besides the field of low-temperature transport, localized many-body states are playing a key role in several research areas as the one of quantum computing\ \cite{roushan2017spectroscopic, laumann2015quantum,altshuler2010anderson}.

The degree of ergodicity of the metal near the many-body localization transition is not clear yet. An ergodic wavefunction roughly means that it has a uniform amplitude in the region of Hilbert space allowed by symmetry constraints\ \ \footnote{ By ergodic we mean a system that obeys the laws of Statistical  Mechanics, which for infinite temperature and using the Eigenstate Thermalization Hypothesis implies that its eigenstates are described by the relevant Gaussian matrix ensemble\ \cite{deutsch2018eigenstate, Rigol2008}.}. Some groups have observed that metallic wavefunctions near this transition are not ergodic\ \cite{Pino15, pino2017,torres2017extended, berkovits2017signature, roy2018multifractality, faoro2018non},  while others support its ergodicity\ \cite{luitz2017ergodic,Luitz2016,Luitz15,vznidarivc2016diffusive}. Given this controversy, it seems sensible to step down and consider simpler models that may exhibit similar behavior than the one expected in the metallic side of the many-body localization transition. Here, we consider one of those simpler models: a single particle hopping in a disordered random-regular graph.

\begin{figure}[t!]
\begin{centering}
\includegraphics[width=0.85\columnwidth]{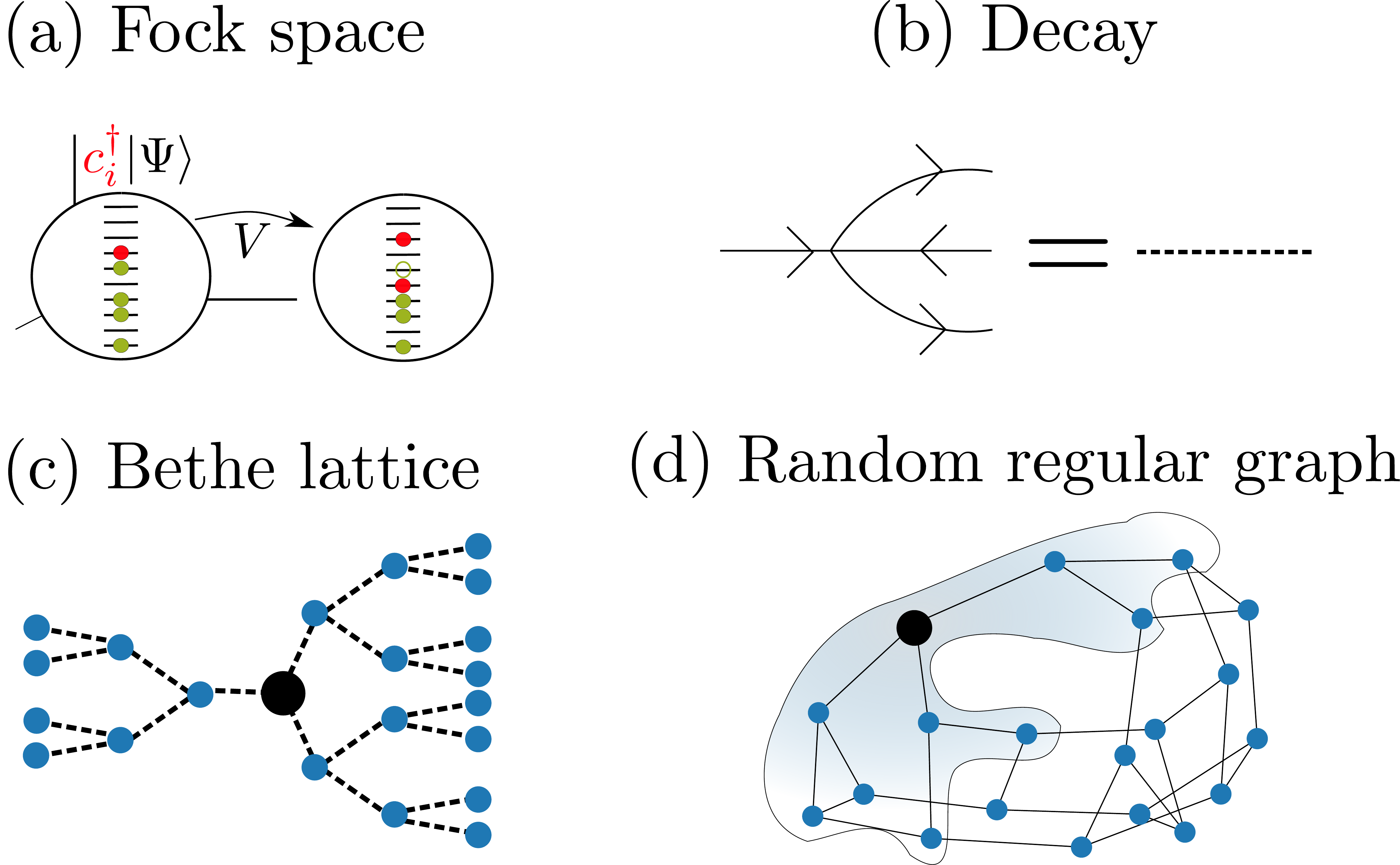}
\par\end{centering}
\vspace{0.1cm} 
\caption{  A particle added to a Fock state of localized single-particles. Due to interactions,  this added particle can decay into two particles and a hole. (b) Representation of this decay process is denoted by a dashed line. The different ways in which the added particle can  successively decay provides with  connections between Fock states\ \cite{Basko2006} similar to the one of a Bethe lattice (c). A random regular graph (d) is locally equivalent to a Bethe lattice, that is, a particle localized in the shaded area around the wide node sees a graph structure similar to the one of a Bethe lattice. 
}\label{Fig1}
\end{figure}

The decay of a local excitation in a many-body and low-dimensional system has similarities to the one of Anderson localization in a Bethe lattice\ \cite{altshuler1997quasiparticle,biroli2020anomalous}, see Fig.\ \ref{Fig1}. The analysis of the later shows that there is a non-ergodic metallic phase where eigenstates are multifractal\ \cite{kravtsov2018non,Biroli2017}. The Rosenberg-Porter model of random matrices\ \cite{rosenzweig1960,altland1997perturbation} may also capture some basic features of non-ergodic many-body metals\ \cite{tarzia2020many}. Despite being proposed  long ago, the existence of those non-ergodic states has been reported recently\ \cite{Kravtsov2015} followed by many other studies\ \cite{facoetti2016non,monthus2017multifractality,truong2016eigenvectors,von2017non,amini2017spread,de2018survival, Bera2018,de2019survival,kravtsov2020localization}. The critical exponent that controls the divergence of correlation length has been found to be $\nu=1$\ \cite{pino2019ergodic}  for both, the localized to non-ergodic metal and the non-ergodic to ergodic metal transitions.

Another relevant model is a particle hopping in a lattice with the connectivity of a random-regular graph, which is locally equivalent to a Bethe lattice (Fig.\ \ref{Fig1}). Several studies predicted a non-ergodic metallic regime in a random-regular graph with branching number $k=2$\ \cite{Deluca2014, Al2016,kravtsov2018non}, while recent ones  have supported the ergodicity of the metal in general random graphs\ \cite{Tikhonov2016b,tikhonov2019,biroli2018delocalization}. In Refs.\ \cite{tikhonov2019, garcia2017scaling,  garc2019}, it is argued that non-ergodic behavior is caused by large finite-size effects due to a correlation volume that diverges exponentially at the metallic side of the transition with critical exponent $\nu_{m}=1/2$, which is different from the critical exponent in the localized side $\nu_l=1$  \cite{Cuevas2001}. In summary, the previous bibliography contains contradictory results regarding the ergodicity of the metallic side of random-regular graphs, although recent studies point to its ergodicity\ \cite{Tikhonov2016b,tikhonov2019,biroli2018delocalization,garcia2017scaling,  garc2019}.

\begin{figure}[t!]
\begin{centering}
\includegraphics[width=0.97\columnwidth]{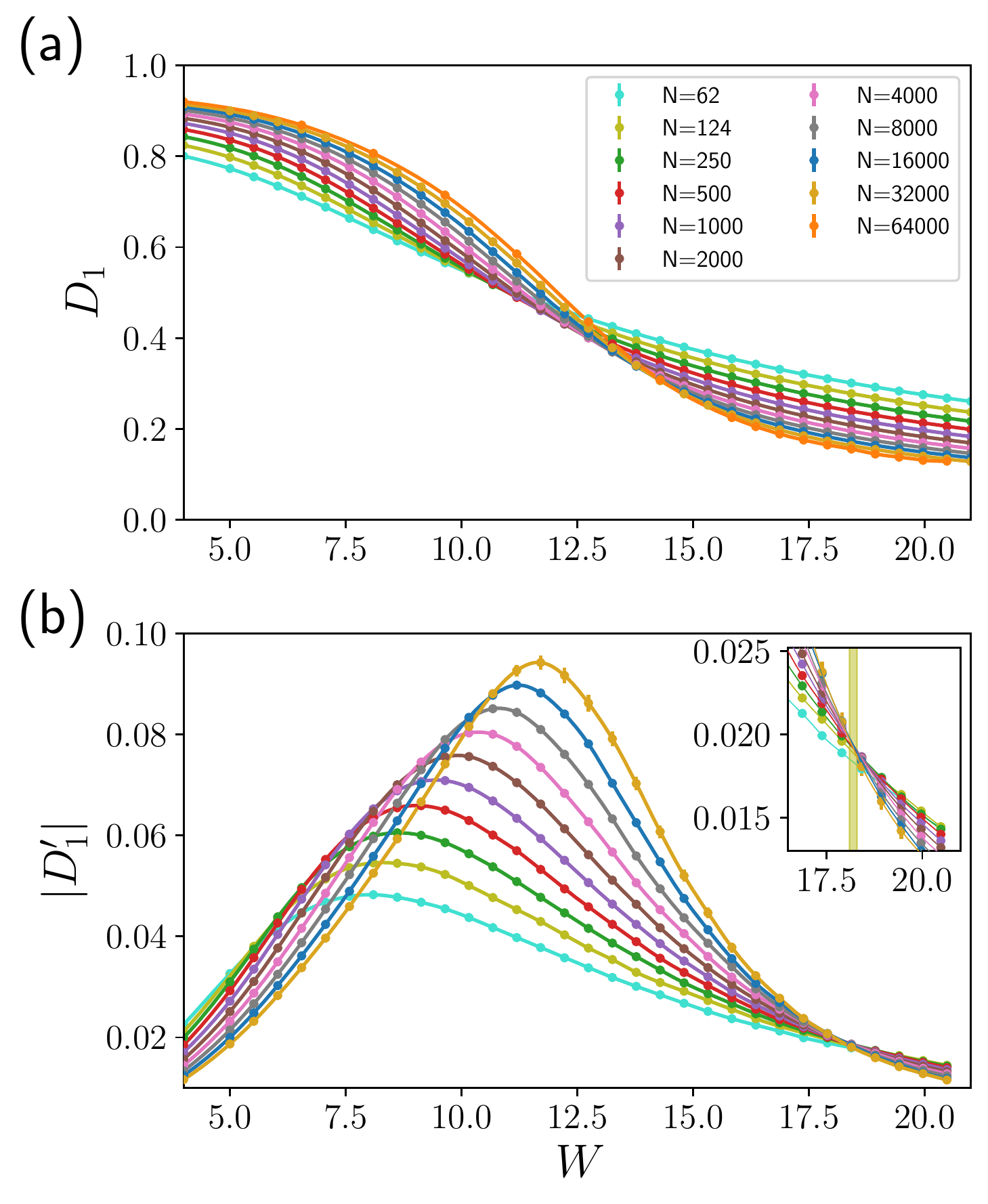}
\par\end{centering}
\vspace{0.1cm}
\caption{ (a) Fractal dimension $D_1=-S/\log(N)$  and (b) its derivative respect disorder $D_1^\prime(N)$, computed with finite difference from the data points of $D_1$, for a random-regular graph with branching number $k=2.$ Data is plotted as a function of disorder for lattices sizes $N=62,124, 250\dots ,32000,64000$ (the points $n=64000$ is not included for $D_1^\prime$). Inset of panel (b):  zoom of the data $D_1^\prime(N)$ in the region $W=16,21.$ The shaded region indicates our estimation of the critical disorder $W_c=18.2\pm 0.1$ from the scaling analysis of $D_1^\prime.$ The solid lines in panels (a) and (b) are spline interpolations of second order for the data of each size.
}\label{Fig:D1}
\end{figure}

Here, we present results that strongly support the non-ergodicity of the metal near the Anderson transition in random regular graph with branching number $k=2.$ We have performed a finite-size analysis for the first two and infinite moments of the wavefunctions amplitudes\  \cite{halsey1986fractal}. Using the scaling hypothesis\ \cite{Fischer1972,rodriguez2010critical}, we are able to accurately determine the critical disorder $W_{\rm cr}=18.2\pm0.09$ and the correlation length critical exponent $\nu=1.00\pm 0.02$ at the Anderson transition, being $\nu$ in agreement with the one expected in a Bethe lattice \cite{kravtsov2018non} and in the Rosenberg-Porter model\ \cite{pino2019ergodic}. The key difference with other numerical studies\ \cite{biroli2018delocalization} is that we analyze the derivative of fractal dimension instead of the fractal dimension itself.

The Hamiltonian for a particle hopping in a random regular graph is ($\hbar =1$) 
\begin{equation}
 H = \sum_{i=1}^N \phi_i c_i^\dagger c_i + \sum_{\av{ij}} c_i^\dagger c_j+c_j^\dagger c_i,
\end{equation}
where $c_i,\ c_i^\dagger$ are fermionic destruction and creation operators.  A realization of this model implies to choose on-site random potentials, we use a box distribution for $\phi_i$ in $[-\frac{W}{2},\frac{W}{2}]$, and a random-regular graph with branching number $k=2$\ \cite{pythonnetwokx}. The second summation in the Hamiltonian runs over the links of that lattice and we denote the number of sites as $N.$ We obtain $20$ eigenstates of the previous Hamiltonian at the center of the band and treat its average as the data for one sample in order to obtain error bars. This realization average is denoted by $\av{\dots}.$ 

\paragraph{Multifractal dimensions.---} We characterize how uniform are the wavefunctions amplitudes using the fractal dimensions $D_{q}= \log_{N}(I_q)/(1-q)$, where the moments are defined as $I_q=\av{\sum_{i=1}^N |\psi_i|^{2q}}$ and $|\psi_i|^2$ is wavefunction amplitude at site $i.$  We are interested on the first two critical dimension $D_1,D_2$ and on  $D_{q\rightarrow \infty}.$ The first one $D_1$ can be computed from the participation entropy  $S=\av{\sum_i |\psi_i|^2\log{|\psi_i|^2}}$, as $S=-D_1 \log(N)+c$. The last one is computed from the maximum wavefunction amplitude $D_\infty=-\log_N(\max{|\psi_i|^2})$, as introduced in Ref.\ \cite{lindinger2019many}. 

The fractal dimensions for an Anderson localized wavefunction are $D_q=0$ for $q\geq 1$, as those wavefunctions have a finite support set. Indeed, the number of sites $N_\xi$  with a finite wavefunction amplitude does not scale with lattice size $N$ but it can be approximated by a geometric serie $N_\xi \approx k^{\xi+1}-1$, being $\xi$ the localization length. On the other hand, we expect that the support set of a metallic eigenfunction scales as a power of the total system size $N^{D}$ with $0<D\leq 1$ (it can be shown that $D=D_1$\ \cite{kravtsov2018non,Deluca2014}). An ergodic state described by a Gaussian Ensemble has vanishingly small wavefunction-amplitude fluctuations around its average value so that $D_q=1$ for all $q.$

We need to determine whether the non-analyticity at the Anderson transition is due to a discontinuity in the $D_q$ or in their derivatives respect disorder $D^\prime_q= \frac{{\rm d}}{{\rm d} W} D_q.$ A direct transition from localized to ergodic states can only occur with a discontinuity in $D_q$ for $q\geq 1$, as in the three-dimensional Anderson model\ \cite{rodriguez2010critical}. On the other hand, the multifractal metallic states near the Anderson transition in Rosenberg-Porter model has a discontinuity in the derivative of fractal dimensions\ \cite{pino2019ergodic}. 

We have plotted $D_1$ in Fig.\ \ref{Fig:D1} (a) as a function of disorder computed as $D_1= S/\log(N)$ for different system sizes. We observe a crossing point which drifts significantly---almost a $50\%$ from the smallest to the largest size---towards larger disorder upon increasing system size as in Ref.\ \cite{biroli2018delocalization}. All those crossing point are very far from the previously reported critical disorder $W_{\rm cr}\approx 18$\ \cite{tikhonov2019,biroli2018delocalization,kravtsov2018non}. We have checked that a large drift of the crossing points also occurs for $D_2,\ D_\infty$, from disorder $W=10$ to $W=14.$  Note that the crossing points in the finite-size data for $D_q$ does not warrant  a direct transition from localization to ergodicity. Indeed, such a crossing point occurs in Rosenberg-Porter model without having such a transition.

\begin{figure}[t!]
\begin{centering}
\includegraphics[width=1.0\columnwidth]{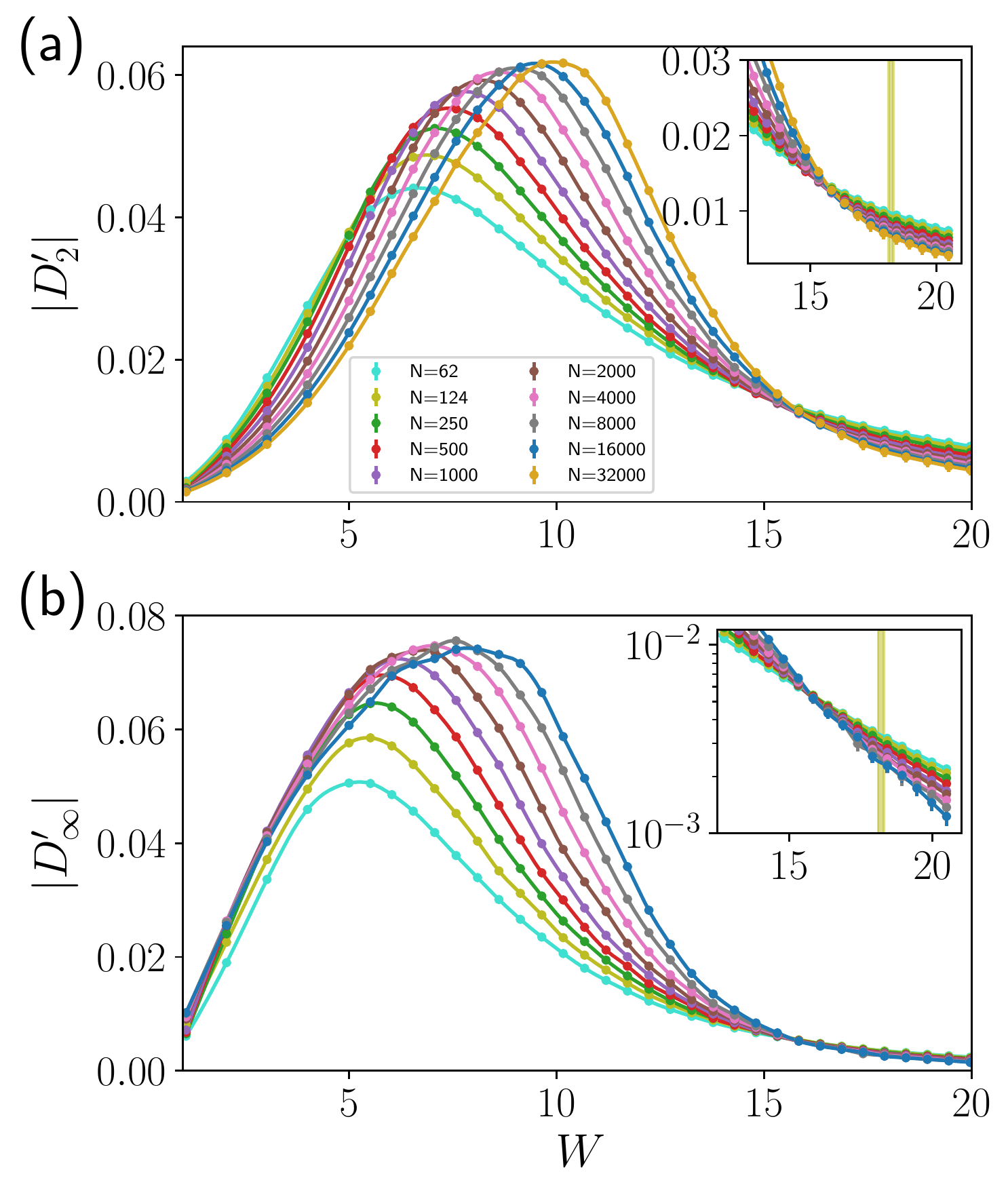}
\par\end{centering}
\vspace{0.1cm}
\caption{ (a) Absolute value of derivative of fractal dimension $D_q$ respect of disorder. The case of second fractal dimension $|D^\prime_2|$ appears in (a) and the case of $|D^{\prime}_{q\rightarrow \infty}|$ as a function of disorder for a random-regular graph with branching number $k=2$ and sizes $N=62,124, 250,500,\dots ,32000.$ The inset of each panel contains a zoom in the region $W=[12.5,21]$, note the semi-log scale in inset of panel (b). The shadow region indicates our estimation of the critical disorder $W_c=18.2\pm 0.1$ from the scaling analysis of $D_1^\prime.$ The solid lines in panels (a) and (b) are spline interpolations of second order for the data of each size.
}\label{Fig:d2_dinfty}
\end{figure}

In the following, we explore the possibility that $D_q=0$ for $q\geq 1$ at the Anderson transition and the discontinuity occurs in its first derivative.  In Fig. \ref{Fig:D1}(b), we have plotted $D_1^\prime(N)$ computed with finite difference from the data in Fig.\ \ref{Fig:D1}(a). This plot reveals a set of crossing point around $W\approx 18$ (inset) which do not significantly shift with increasing lattice size. Crucially, the critical value of $(D_1^\prime)_{\rm cr}$ defined by the crossing points does not show any tendency to diverge. From data in Fig. \ref{Fig:d2_dinfty}, we see that the same picture holds for $D_2^\prime$ and $D_\infty^\prime$, with a crossing point around $W\approx 17.$ Those crossing points shift towards larger disorders when increasing system sizes,  see inset of panel (a) and (b) of Fig.\ \ref{Fig:d2_dinfty}.  

In summary, our data do not support the divergence of fractal dimensions $D_q^\prime$ for $q\geq 1$ at the Anderson transition, which is strictly needed in the case of a transition from localization to ergodicity. Instead, they indicate a discontinuity in the derivatives $D_q^\prime$, as in the Rosenberg-Porter model\ \cite{pino2019ergodic}. The absence of such a divergence  implies the non-ergodicity of the metallic side of random-regular graphs with $k=2.$

\paragraph{Scaling and critical properties} In the following, we use the scaling hypothesis to find the critical properties at the Anderson transition. Near a second order phase transition, the singular part of a quantity can be expressed as $A= \xi^{\zeta/\nu} g(L/\xi)$, where $L,\ \xi$ are system size and correlation length\ \cite{Fischer1972}. In the thermodynamic limit, we have $\xi\sim |W-W_{\rm cr}|^{-\nu}$ and $A\sim|W-W_{\rm cr}|^{-\zeta}.$ The scaling function controls the crossover at finite sizes, $g(x)\sim x^{\zeta/\nu}$ when $x=L/\xi\ll 1$ and $g(x)$ is constant when $x=L/\xi\gg 1.$ For practical purposes is better to work with a scaling function $f(x)=g(x^\nu)/x^\zeta$, so that $f$ can be expanded as a polynomial near the critical point. Taking into account irrelevant corrections to scaling, we have\ \cite{cardy1996scaling,rodriguez2010critical,slevin1999corrections}
\begin{equation}\label{eq:scaling}
A=L^{\zeta/\nu}f\left(\rho L^{1/\nu},\eta L^{-y}\right). 
\end{equation}
Near the critical point $\rho=(W-W_{\rm cr})$ and for irrelevant corrections $y>0.$ In a random-regular graph, the finite-size crossover is expected when the correlation length is similar to the number of generations $\xi\sim \log(N).$ Thus, we replace $L=\log(N)$ in the scaling law Eq.\ (\ref{eq:scaling}). The basic idea behind finite-size scaling is that choosing the right critical parameters $W_{\rm cr},\ \nu$ and subtracting irrelevant corrections, one should be able to collapse all the curves for $A$ when plotted as a function of the scaling variable $x=L^{1/\nu}(W-W_{\rm cr}).$  We note that the drift of the crossing points for $D_q^\prime$ near $W\approx 18$  in Figs.\ (\ref{Fig:D1}, \ref{Fig:d2_dinfty}) indicates that corrections to scaling are important. 

\begin{figure}[t!]
\begin{centering}
\includegraphics[width=1.03\columnwidth]{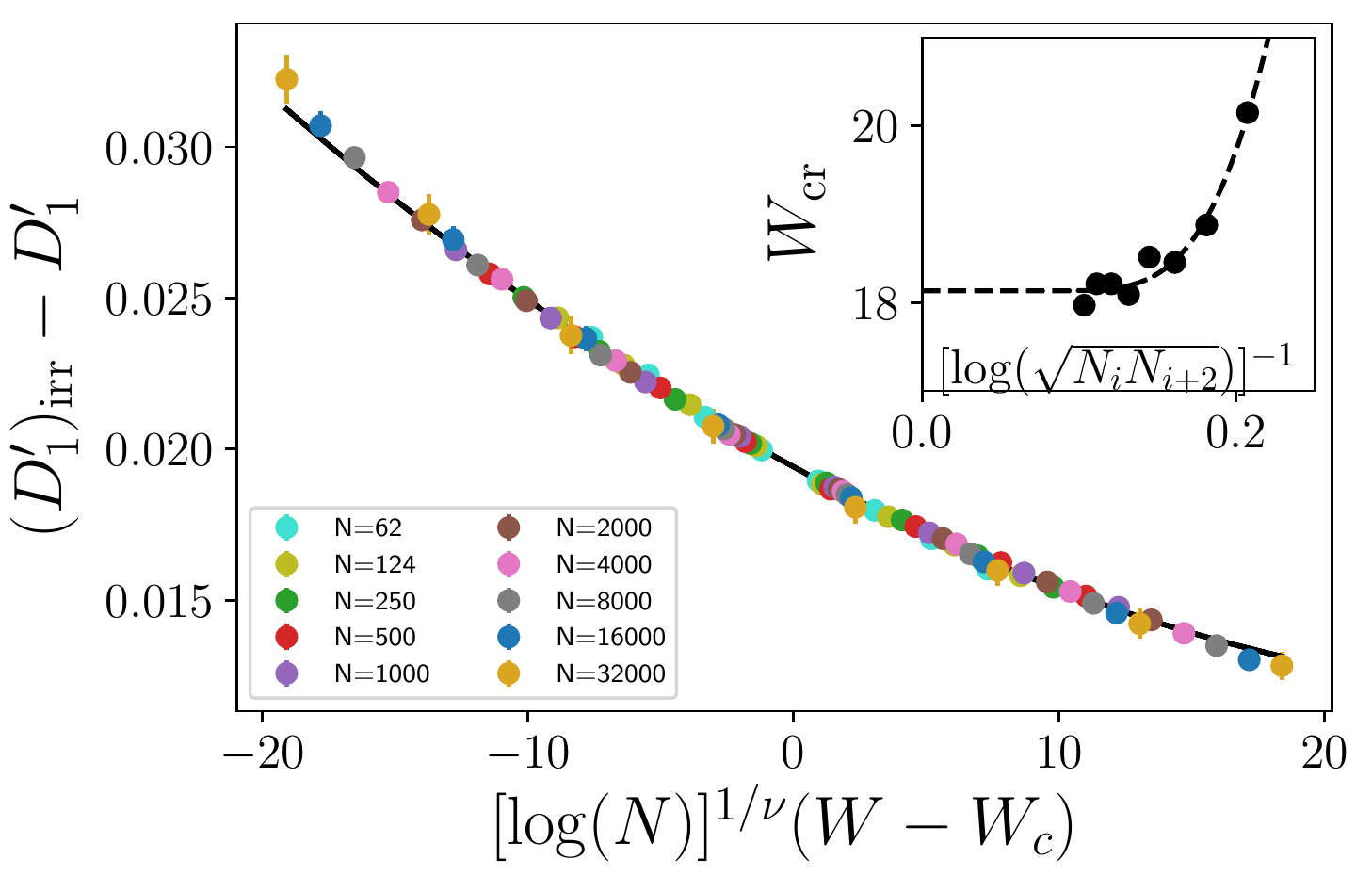}
\par\end{centering}
\vspace{0.1cm}
\caption{Best data collapse for the derivative of first fractal dimension $D_1^{\prime}(N)=S^{\prime}/\log(N)$ for disorder $W=[16, 20]$ and sizes $N=62,124,250,\dots 32000.$ We use a scaling function of second order and irrelevant corrections $D_1^{\prime}= f(x) (1+\eta\  [\log(N)]^{-y})$, which gives critical parameters $\nu =1.00\pm 0.02$, $W_{\rm cr} =18.2\pm 0.1$, $(D_1^\prime)_{\rm cr}=(-1.94\pm 0.08)10^{-2}$ and irrelevant exponent $y=5\pm 2.$ We employ a second order polynomial for the function $f,$ (solid line) which gives a reduced-$\chi^2$ of $0.66.$ The main panel contains $(D_1^\prime)_{\rm irr}-D_1^\prime$ being the irrelevant corrections $(D_1^\prime)_{\rm irr}=f(x)\ \eta\  [\log(N)]^{-y}.$ Inset contains the critical disorder $W_{\rm cr}$, defined as the disorder at which $D_1^{\prime}(N_i)$ and $D_1^{\prime}(N_{i+2})$  crosses, as a function of the inverse of logarithm of the geometric mean of the sizes $L_i$ and $L_{i+2}.$ The dashed line is a fit  to the law $W_{\rm cr}+\eta [\log(N)]^{-y}$ which gives $W_{\rm cr}=18.13\pm 0.08$ and $y=6\pm 2$ in agreement with results from the scaling in the main plot.
}\label{Fig:scaling}
\end{figure}

In Fig.\ref{Fig:scaling}, we show the result of a scaling analysis of $D_1^\prime$ based on Eq.\ \ref{eq:scaling} with $\zeta =0$, as we have previously argued that our data do not support the divergence of $D_1^\prime$ at criticality. We choose $D_1^\prime$ because this quantity has much smaller corrections to scaling that $D_2^\prime$ or $D_\infty^\prime$. We use a small disorder interval $W=[16,20]$  so that we can use a first order expansion in the fields $\rho$ and $\eta$ of Eq.\ \ref{eq:scaling}. We employ irrelevant corrections of the form $f= \bar{f}(x)(1+\eta [\log(N)]^{-y})$, because it gives good quality fittings and the result for $y$ is consistent with the scaling analysis of the drift of crossing points $W_{\rm cr}$ at criticality, see inset of Fig.\ \ref{Fig:scaling}. The function $\bar{f}$ is approximated with a second order polynomial, which gives the closest to one reduced-chi square  $\chi_r^2=0.66$, and we checked that bootstrap techniques gives similar error estimation\ \cite{andrae2010}. 

The best data collapse for $D_1^\prime(N)$ gives $\nu=1.00\pm 0.02$, $W_c=18.2\pm 0.1$ with irrelevant exponent $y=5\pm 2.$ The estimation of the critical point is in good agreement with the ones in Refs.\ \cite{tikhonov2019,biroli2018delocalization} and not very far from others\ \cite{kravtsov2018non,abou1973selfconsistent}. Our result for $\nu$ agrees with the theory of \cite{kravtsov2018non} but disagree with Refs.\ \cite{tikhonov2019,garc2019,garcia2017scaling}, where $\nu=1/2$ in the metallic side. We note that a fitting without irrelevant corrections but with the data for $N>500$ gives an estimation $W_{\rm cr}=18.27\pm 0.07$ and $\nu=0.93\pm0.06$, but the quality of the fitting is worst than the previous one.

\begin{figure}[t!]
\begin{centering}
\includegraphics[width=.990\columnwidth]{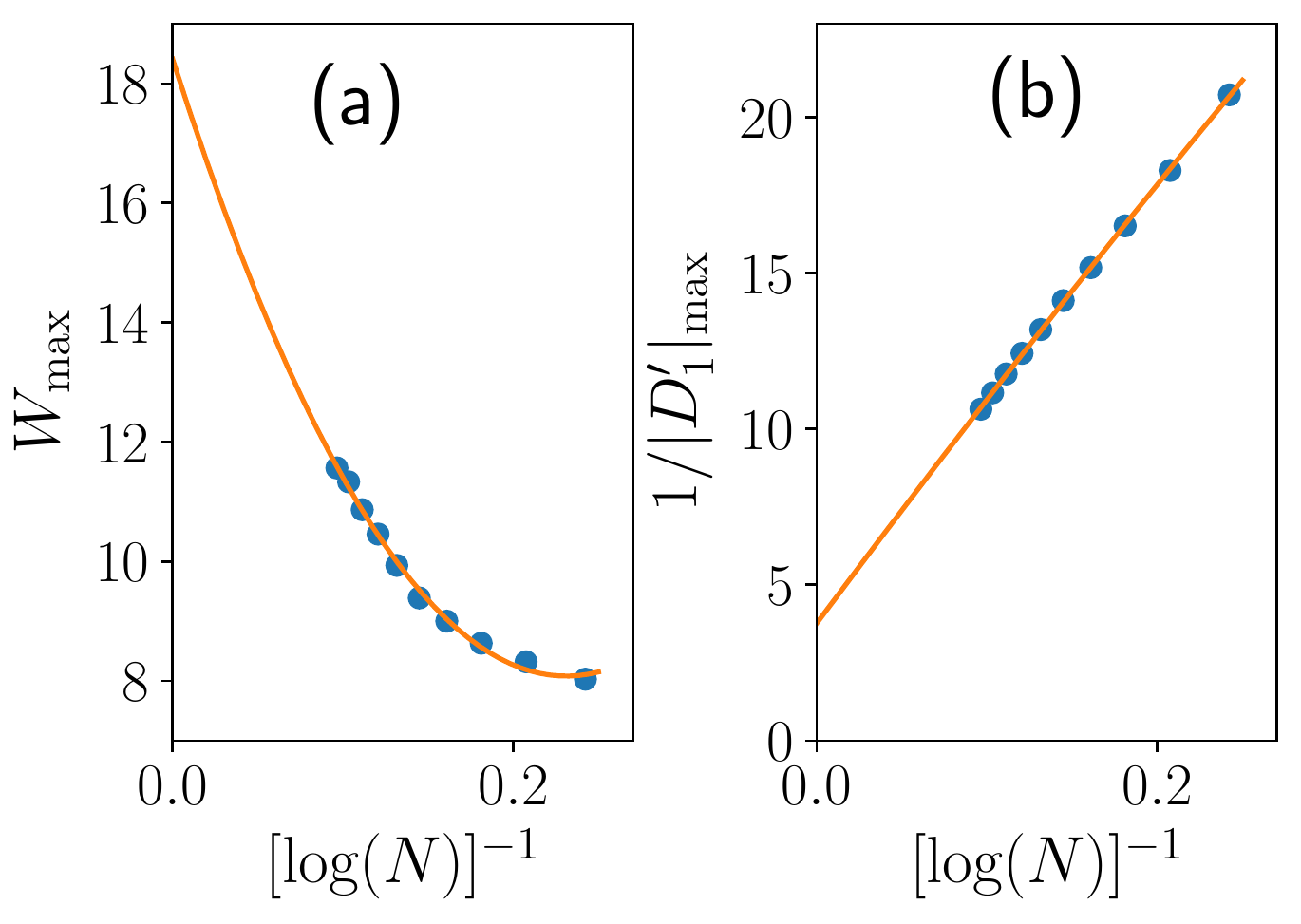}
\par\end{centering}
\vspace{0.1cm}
\caption{  The disorder at which the maximum of $|D_1^{\prime}|$ is reached (a) and  the inverse of the maximum value $1/|D_1^{\prime}|$ (b) as a function of the inverse of the logarithm of the system size. The straight lines in panel (a) and (b) are second and first order polynomial fittings, respectively. They allow to extrapolate the position of the maximum to $W_{\rm max}=18\pm 4$ and of maximum to $1/|D_1^\prime|_{\rm max}= 4\pm 0.3$ in the thermodynamic limit. Error bars accounts for the differences in the extrapolated values when using a polynomial of one degree higher in each case. 
}\label{Fig:ext}
\end{figure}

\paragraph{Metallic side of the Anderson transition---} 
Our previous results strongly support that fractal dimensions derivatives $D_q^\prime$ do not diverge at the Anderson transition and, thus, the metallic side of this transition must be non-ergodic as $D_q<1$.
Furthermore, our numerical data does not support the existence of additional phase transition, different from the Anderson one, between the non-ergodic and ergodic metals. The case of a first order transition, as explained in Ref. \cite{kravtsov2018non},  would imply a divergence of $D_q^\prime$ in the thermodynamic limit. The absence of such a divergence in $D_\infty^\prime$ is clear from the data in Fig.\ \ref{Fig:d2_dinfty} (b), as the maximum  is approximately constant $|D_\infty^\prime|_{\rm max}\approx 0.07$ for sizes $N>500$. We can draw a similar conclusion from the $D_1^\prime$ data. To see this, we have extrapolated to the thermodynamic limit the maximum value, and its location, of fractal dimension derivative $D_1^\prime$ appearing in Fig.\ref{Fig:D1}(b). In Fig.\ \ref{Fig:ext} panel (a) and (b), we can see the location of those maxima and their values, respectively, as a function of $1/\log(N)$. We obtain an extrapolated value at thermodynamic limit $1/|D_1^\prime|_m=4.0\pm 0.3$, being the extrapolated location close to the Anderson critical point, see Fig.\ \ref{Fig:ext}. Our data thus indicate that the derivative of fractal dimensions, $D_1^\prime$ for $q\geq 1$ remains finite even at the maximum observed in Fig\ \ref{Fig:D1}(b).

The possibility of a second order non-ergodic to ergodic transition, with a discontinuity in $D_q^\prime$ as in the Rosenberg-Porter model\ \cite{pino2019ergodic}, is not supported by our numerical data neither. We note that all the plots for derivative of fractal dimensions, Fig.\ \ref{Fig:D1} (b) and panels (a) and (b) of Fig.\ \ref{Fig:d2_dinfty}, contains two set of crossing points where the curves for consecutive sizes crosses. The crossing points in the first set appear near $W\approx 18$ and they can be ascribed to the Anderson transition. The other  crossings occur around $W<13$,  drifting upon an increase of system size. This last set of crossing point are not compatible with a second order phase transition because the slopes of the curves at the crossing points decrease. This is clearly seen for the derivative of $D_2^\prime$ for the crossings at $W\approx 8$ and from the $D_\infty^\prime$ data in Fig. \ref{Fig:d2_dinfty}.

\paragraph{Conclusions.---}

 We have shown that fractal dimensions derivatives $D_1^\prime, D_2^\prime$ and $D_\infty^\prime$ do not diverge at the Anderson critical point for a random-regular graph of branching number $k=2.$ This strongly supports that ergodicity is not restored at the metallic side of the Anderson transition, so there is a finite region of multifractal metallic states. We have performed a finite size scaling of the first fractal dimension derivative---which showed the smallest finite-size corrections to scaling---obtaining critical disrder $W_{\rm cr}=18.20\pm 0.09$ and $\nu=1.00\pm 0.02$, with large irrelevant exponent $y\approx 5.$ We have further discussed that our data does not support the existence of an additional non-ergodic  to ergodic metal transition, so there is a crossover from non-ergodic to ergodic behavior. 
 
\begin{acknowledgments}
\paragraph{Acknowledgments.---} I thank A. Rodr\'iguez, P. Serna and J. J. Garc\'ia-Ripoll for useful discussion and comments. Financial support by Fundación General CISC (Programa Comfuturo) is acknowledged. The numerical computations have been performed in the cluster Trueno of the CSIC.
\end{acknowledgments}

\bibliography{./MBLandNonequilibrium}
\end{document}